\documentclass[prb,twocolumn,showpacs,amsmath,amssymb]{revtex4-1}

\usepackage{graphicx}
\usepackage{epstopdf} 
\usepackage{dcolumn}
\usepackage{bm}
\usepackage{hyperref}
\usepackage[mathlines]{lineno}
\usepackage[usenames,dvipsnames,svgnames,table]{xcolor}
\usepackage{times}
\usepackage{color}
\newcommand{\etal}{\textit{et al.}}
\newcommand{\beq}{\begin{equation}}
\newcommand{\eeq}{\end{equation}}
\newcommand{\beqn}{\begin{eqnarray}}
\newcommand{\eeqn}{\end{eqnarray}}
\newcommand{\he}[1]{\textsuperscript{#1}He}

\begin{document}
\title{Superfluid nanomechanical resonator for quantum nanofluidics}

\author{X. Rojas}
\email{xavier@ualberta.ca}
\author{J.P. Davis}
\email{jdavis@ualberta.ca}
\affiliation{Department of Physics, University of Alberta, Edmonton, Alberta, Canada T6G 2E9}
\date{\today}

\begin{abstract}
We have developed a nanomechanical resonator, for which the motional degree of freedom is a superfluid \he4 oscillating flow confined to precisely defined nanofluidic channels. It is composed of an in-cavity capacitor measuring the dielectric constant, which is coupled to a superfluid Helmholtz resonance within nanoscale channels, and it enables sensitive detection of nanofluidic quantum flow. We present a model to interpret the dynamics of our superfluid nanomechanical resonator, and we show how it can be used for probing confined geometry effects on thermodynamic functions. We report isobaric measurements of the superfluid fraction in liquid \he4 at various pressures, and the onset of quantum turbulence in restricted geometry.
\end{abstract}


\maketitle


\section{\label{sec:level1_int}Introduction}
At low temperatures, liquid \he3 and \he4 transition into superfluids, which exhibit exotic properties - such as dissipationless flow - as a result of macroscopic quantum coherence. The coherent motion of the superfluid state is described by an order parameter, whose spatial fluctuations are correlated over a length scale given by the coherence length, $\xi$. This coherence length diverges at the superfluid transition and reaches a finite value in the low temperature limit ($\xi_0=20 - 80$ nm in \he3~\cite{Wheatley1975} and $\xi_0=0.1$ nm in \he4~\cite{Sukhatme2001}). By confining these quantum fluids in well-defined structures of size comparable to the coherence length, non-bulk phenomena can be revealed. For instance, nanofluidic confinement has allowed the study of the order parameter fluctuation spectrum~\cite{Gasparini2008}, as well as proximity coupling~\cite{Perron2012}, near the superfluid transition of \he4. In \he3, confinement approaching the coherence length is predicted to result in new superfluid phases~\cite{Vorontsov2007,Levitin2013} due to geometrically induced distortion of the order parameter. This distortion is directly related to surface states, which are predicted to be Majorana fermions at the surface of \he3-B~\cite{Salomaa1988,Chung2009,Park2014,Mizushima2014}.

Despite the fact that these superfluids are well studied, only a few experiments are capable of measuring such tiny volumes, or small surface effects at the edge of bulk volumes. The most powerful experimental techniques to date are nuclear magnetic resonance NMR~\cite{Levitin2013} in \he3 and heat capacity~\cite{Gasparini2008} in \he4, although new techniques using nanomechanical structures are promising~\cite{Gonzalez2013,Gonzalez2013b,Defoort2014}. Yet numerous theoretical predictions go untested because the right experimental probes do not exist.  For example, a signature of Majorana fermions in superfluid \he3-B confined to channels of order $\xi_0$ has been predicted in the superfluid density, $\rho_s/\rho$~\cite{Wu2013}.  Previous studies of superfluid density by studying mass flow in confined superfluids have been limited to Helmholtz resonators in large arrays of particle-etch track pores~\cite{Brooks1979}, stacks of thousands of slabs of superfluid~\cite{Freeman1988, Freeman1990}, or hindered by normal state decoupling.\cite{Dimov2010}  In this work, we demonstrate a quantum nanofluidic experiment capable of measuring both the static properties of a highly confined superfluid - in particular the dielectric constant and total density - as well as dynamic properties, namely mass flow, superfluid density, and dissipation; and we use this quantum nanofluidic device to study superfluid \he4, probing just nanoliters of liquid with high sensitivity.  One result of our system is a measurement of dissipation in thin channels ($\sim500$ nm) at velocities into the quantum turbulence regime~\cite{Vinen2002,Bradley2011,Schmoranzer2014,Ahlstrom2014}. This presents a scenario in which vortex lines may be pinned by surfaces in the confined geometry, and it should open the door to new theoretical and experimental studies.

\begin{figure}[b]
\centering
\includegraphics[width=8.6cm]{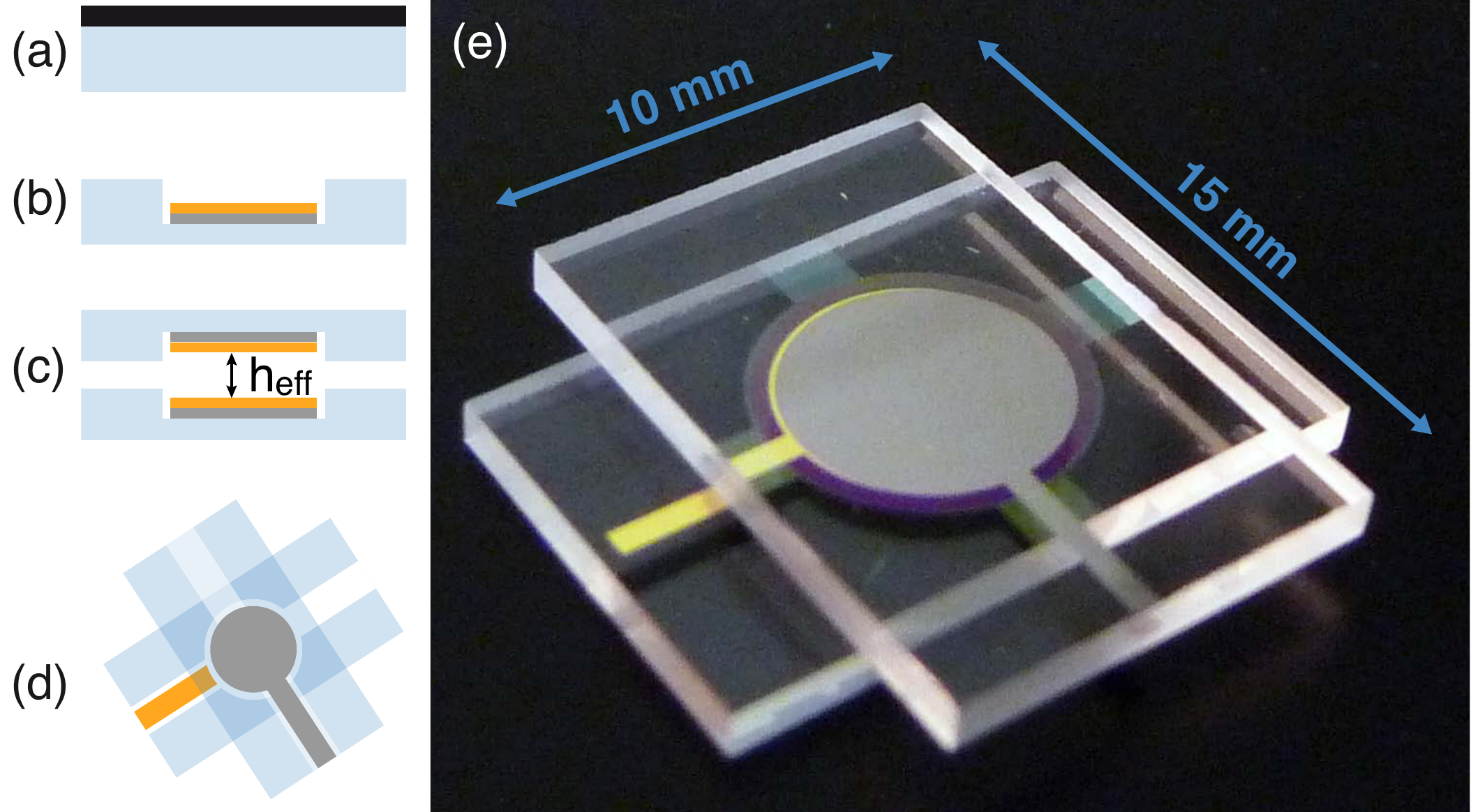}
\caption{The superfluid nanomechanical resonator is defined by an oscillating mass of superfluid \he4 confined in the channels of a nanoscale structure. (a - d) The nanofabrication process for our devices (see appendix A for details). (e) Photograph of the completed device with an effective cavity height $h_{\rm eff}\simeq900$ nm and a channel height $h_{\rm cha}\simeq550$ nm. The light blue, yellow and purple colors are the result of optical interference.}
\label{Fig_1}
\end{figure}

In the dynamic regime, our experiment is a superfluid Helmholtz resonator~\cite{Backhaus1997a,Backhaus1997b} and its behavior is well described by analogy with a nanomechanical mass-on-a-spring.  Here the mass is given by the amount of superfluid within a volume of 9 nL, a temperature dependent quantity, which ranges from $\approx24$ ng ($\rho_s/\rho=0.02\%$) at $T-T_{\lambda}\simeq2$ mK to $\approx960$ ng ($\rho_s/\rho=80\%$) near 1.6 K. This represents an unusual nanomechanical system with small moving mass and intrinsic quantum properties, and it may also provide an opportunity to study mechanical resonators in the quantum regime. Unlike classical mechanical resonators, superfluids are dissipationless coherent macroscopic quantum states. At very low temperature where the normal component is negligible, the mechanical quality factor of a superfluid resonator can be exceedingly large ($Q>10^{10}$), as in the ground-breaking work of De Lorenzo~\etal~\cite{DeLorenzo2014}, in which they measured $Q=10^6$ at 10 mK. Here, the quality factor of our superfluid nanomechanical resonator increases by three orders of magnitude between $T_\lambda$ and 0.7 K.

\section{\label{sec:level1_exp}Experiment}
In the experiment, we have immersed the nanofabricated structure shown in Fig.~\ref{Fig_1}e in a liquid \he4 bath sealed in a copper sample cell mounted on a cryostat. The device consists of features etched into glass, where the etch height defines the relevant confinement length~\cite{Duh2012,Rojas2014}.  Specifically, we form a circular cavity (radius $r_{\rm cav}=2.5$ mm and height $h_{\rm cav}= 1100$ nm) and four channels (length $l_{\rm cha}=2.5$ mm, width $w_{\rm cha}=1.6$ mm and height $h_{\rm cha}=550$ nm).  Electrodes (height $h_{\rm ele}=100$ nm and radius $r_{\rm ele}=2$ mm) were deposited on both inner sides of the cavity before bonding the device, and they are electrically contacted from two of the channels. The effective confinement length in the cavity is given by the distance between the two electrodes ($h_{\rm eff}=h_{\rm cav}-2h_{\rm ele}=900$ nm). 

The completed device realizes a parallel plate capacitor with a nanoscale gap, which can be used to study the dielectric properties of a fluid in the gap via measurement of the capacitance. Specifically, $C=(A_{\rm ele}\epsilon_0\epsilon_r)/(h_{\rm eff})$, with $h_{\rm eff}$ the effective confinement length, $A_{\rm ele}=\pi r_{\rm ele}^2$ the surface area of the electrodes, $\epsilon_0$ the vacuum permittivity and $\epsilon_r$ the relative permittivity of the liquid. 

\section{\label{sec:level1}Results and Discussion}

\begin{figure}[t]
\centering
\includegraphics[width=8.6cm]{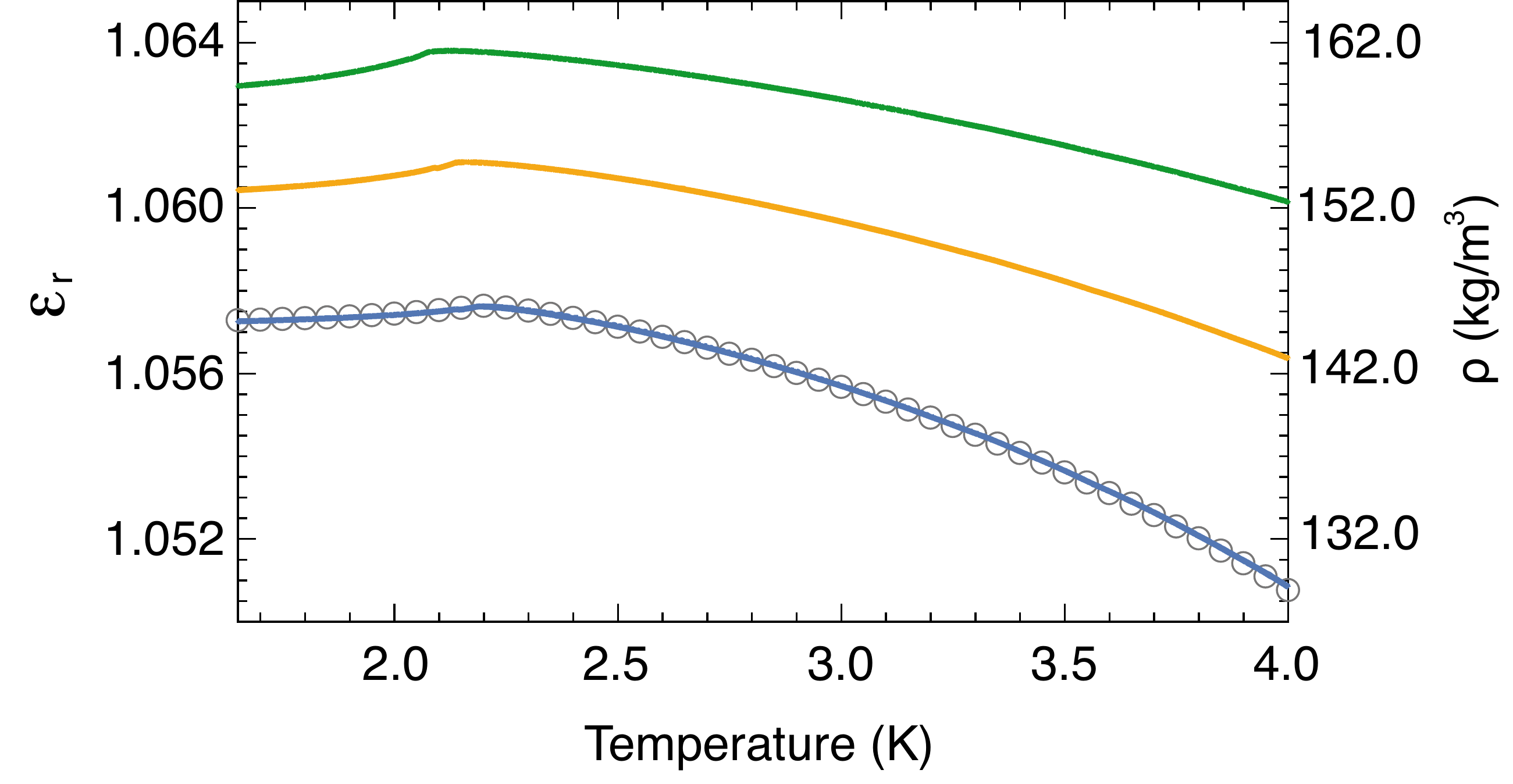}
\caption{Measurements of the dielectric constant $\epsilon_r$ of $11$ nanoliters of liquid \he4 confined between the electrodes of the nanofluidic cavity, from 4 to 1.65 K at saturated vapor pressure (blue), 5 bar (orange) and 10 bar (green). Circles are bulk values at saturated vapor pressure from Ref.~\citenum{Donnelly1998}. The lambda transition occurs at the kink in $\epsilon_r$.}
\label{Fig_2}
\end{figure}

We have used this device to measure the dielectric constant of liquid \he4 confined between the electrodes (Fig.~\ref{Fig_2}). The dielectric constant is given by $\epsilon_r=(C-C_s)/(C_0-C_s)$, with $C_0\simeq117$ pF the capacitance of the empty capacitor and $C_s$ the stray capacitance originating from the capacitance of our measurement coax. We computed the stray capacitance, $C_s=1.0$ pF for our setup, by fitting our data at saturated vapor pressure with the data of Donnelly~\etal~\cite{Donnelly1998}  We see excellent agreement with the temperature dependence of the bulk values for the dielectric constant, shown in Fig.~2, despite the fact that we are probing just 11 nL of liquid. 

By measuring the relative dielectric constant, $\epsilon_r$, one can obtain the density, $\rho$, via the Clausius-Mossotti relation, which works well for a non-polar liquid such as \he4~\cite{Chan1977}. Measurement of the temperature dependent density is then a characterization step of our nanomechanical resonator, analogous to measuring the mass of the mass-on-a-spring.  Results are shown on the right hand side of Fig.~2. Such dielectric measurements may also be relevant in superfluid \he3, since the electric field couples weakly to the order parameter of \he3~\cite{Volhardt1990,Swift1980}, and it is currently unknown whether there are electric field effects in \he3 that could be probed with this technique.  

Beyond static measurements of the fluid density, one can use these devices to perform dynamic quantum fluid flow experiments. Indeed, a nonzero flow in the channels induces a density change in the cavity, which can be detected as a change in the relative dielectric constant. The flow of liquid \he4 is well described by the two-fluid model~\cite{Wilks1967} composed of a normal component $\rho_n$ and a superfluid component $\rho_s$.  Below a critical velocity, the two fluids behave independently with their own local mass flow velocities $v_n$ and $v_s$, and the viscosity of the fluid is entirely given by the viscosity of the normal component $\eta_n$. In a confined geometry, the normal component can be clamped by the walls in the dynamic regime if the confinement length is small compared to the viscous penetration depth~\cite{Lauter1979}. The viscous penetration depth is defined as $\lambda_\nu=\sqrt{(2\eta_n)/(\rho_n\omega)}$, with $\omega / 2\pi$ the frequency of the oscillating flow. In this experiment, the frequency did not exceed 5 kHz, and therefore $\lambda_\nu > 1.0\ \mu$m. This is roughly four times the distance to a wall in our device, therefore the normal component is mostly clamped and only the superfluid oscillates in the channels.   The dynamic resonance described below, therefore, only appears in the superfluid phase ($T<T_\lambda$) when $\rho_s/\rho\neq0$.  In the case of dc flow, however, it is possible to measure a contribution from the normal density component even for nanoscale channels~\cite{Savard2009,Savard2011}.

\begin{figure}[b]
\centering
\includegraphics[width=8.6cm]{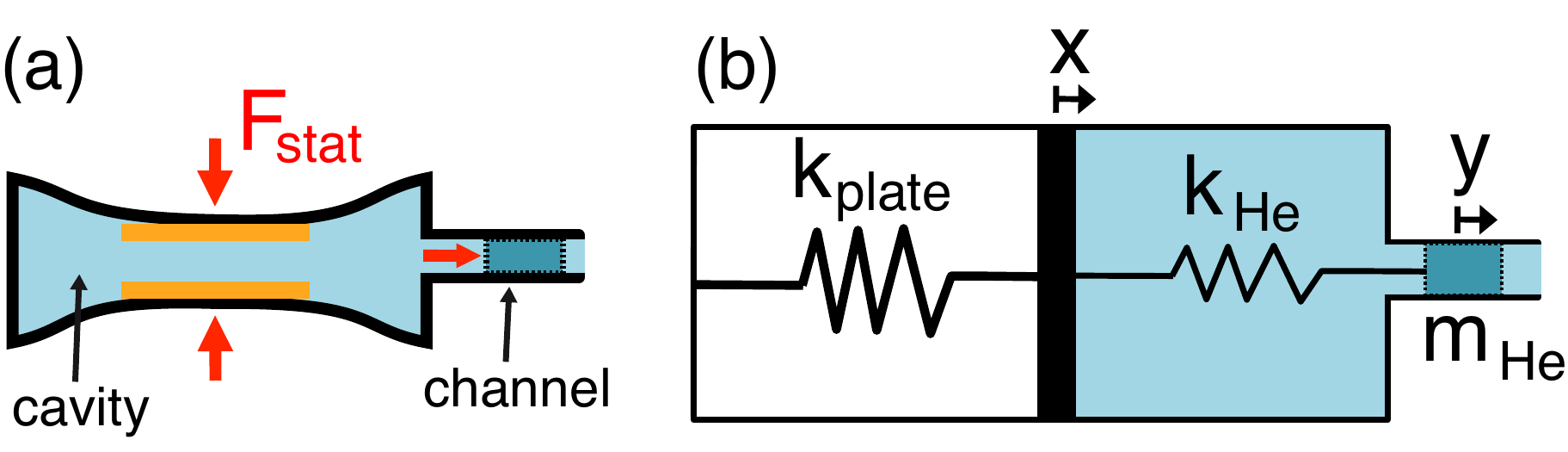}
\caption{Simplified schematic of the superfluid nanomechanical resonator. (a) The electrostatic force, $F_{\rm stat}$, deforms the glass cavity and generates a pressure gradient across the channel. (b) A schematic of an equivalent  mass-spring system - the plate moves by $x$ and the mass of superfluid, $m_{\rm He}$, responds by moving a distance $y$.}
\label{Fig_3}
\end{figure}

We drive an ac Helmholtz resonance~\cite{Backhaus1997a,Backhaus1997b} with the same voltage $V_d$ that is used to measure the capacitance of the nanofluidic capacitor. The electric field $E_d=V_d/h_{\rm eff}$ between the electrodes produces an attractive electrostatic force between the glass plates, $F_{\rm stat}=(1/2)A_{\rm ele}\epsilon_0\epsilon_rE_d^2$, which bend under this load (Fig.~\ref{Fig_3}). This deformation produces a pressure increase in the cavity, which induces a flow in the channels.  It is interesting to note that other forces such as the electrostrictive and the Casimir forces, while of smaller magnitude, exist in the system and can be enhanced by modifying the present geometry (see Appendix G).  

We describe the electrostatically driven Helmholtz resonance in the channels~\cite{Backhaus1997a, Backhaus1997b}, with a mass-on-a-spring model (see Fig.~\ref{Fig_3}). In this model, we consider a superfluid mass $m_{\rm He}=4\rho_s l_{\rm eff} a$, with $l_{\rm eff}$ the effective length of the channel, which accounts for the end effects and is slightly larger than its physical length, and $a=w_{\rm cha}h_{\rm cha}$ is the cross-sectional area of the channel. This mass is attached to a spring of stiffness $k_{\rm He}$, accounting for the bulk modulus of the liquid in the cavity, which is attached in series to another spring of stiffness $k_{\rm plate}$, accounting for the flexural rigidity of the glass plates. The effective stiffness of this resonator, $k_{\rm eff}$, is a combination of these two springs in series, multiplied by a geometric factor. The resonance frequency of the mechanical system is then given by
\beq\label{Eq:resonance}
\omega_0^2= \frac{k_\textrm{eff}}{m_{\rm He}}=4\left(\frac{\rho_s}{\rho}\right)\left(\frac{a}{A_{\rm plate}^2 l_{\rm eff}\rho}\right)\frac{k_{\rm plate}}{1+\Sigma}
\eeq
with $\Sigma=k_{\rm plate}/k_{\rm He}$, $k_{\rm He}=(A_{\rm plate}^2)/(\chi V_{\rm cav})$, $V_{\rm cav}$ is the volume of the cavity, $\chi$ is the compressibility of liquid \he4, and $k_{\rm plate}=1.94\times10^7$ N/m is the bending stiffness of the glass plates, which we measured.  Derivation and details can be found in Appendix E.

The oscillating superfluid in the channels generates a density oscillation in the cavity that can be measured as a time-dependent dielectric constant. We measured this by studying the ac response of the dielectric constant of the superfluid using a frequency-dependent capacitance measurement. The electrodes of the nanofluidic capacitor were connected to a capacitance bridge (General Radio 1615-A). The drive voltage $V_{d}$ applied to the capacitor was produced by a function generator (Stanford Research DS345), and the response of the capacitance bridge was amplified and measured with a lock-in amplifier (Stanford Research SR830) synchronized to the function generator. Measurement of the two quadratures allowed us to extract the real (in-phase) and imaginary (out-of-phase) parts of the complex dielectric constant ($\epsilon_r=\epsilon_r^{\prime}-\mathrm{i}\epsilon_r^{\prime\prime}$), which were simultaneously fit to a damped harmonic-oscillator equation,
\beq\label{eq:harm_osc}
\ddot{\epsilon_r}+\frac{\omega_0^2}{Q}\dot{\epsilon_r}+\omega_0^2\epsilon_r=\omega_0^2\epsilon_r^{d}
\eeq
with $Q$ the quality factor and $\epsilon_r^d$ the driving term proportional to the force applied to the plates, $F_{\rm stat}$.

\begin{figure}[t]
\centering
\includegraphics[width=8.6cm]{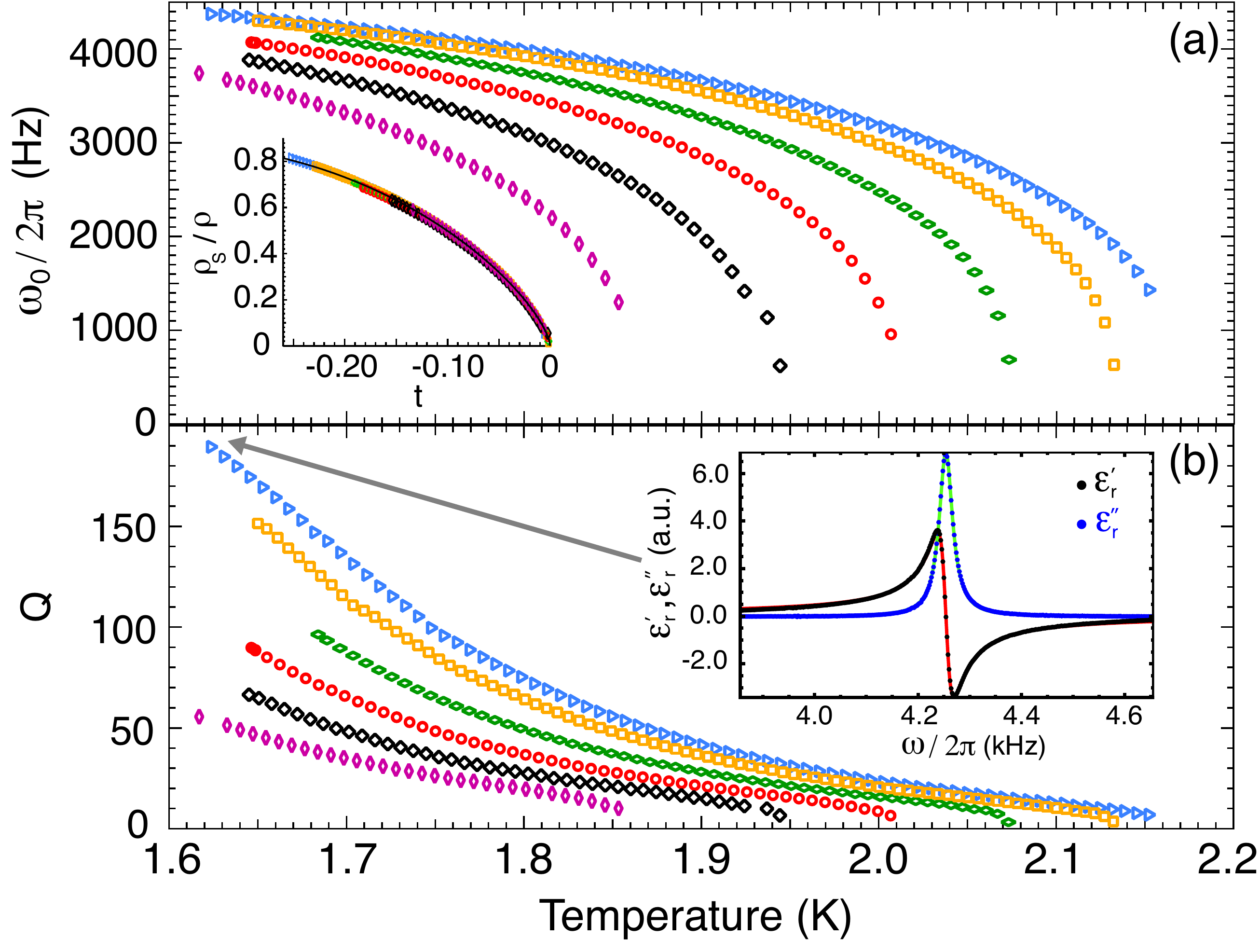}
\caption{Temperature dependence of the superfluid nanomechanical resonator  (a) frequency and (b) quality factor, taken at a drive voltage $V_d=5$ V, for various pressures: 2 (blue triangles), 5 (orange squares), 10 (green diamonds), 15 (red circles), 20 (black diamonds) and 25 (purple diamonds) bar. The inset of (a) is the superfluid fraction, $\rho_s/\rho$, extracted from the resonance frequency using Eq.~\ref{Eq:resonance} with $T_\lambda$ taken from Maynard sound measurements~\cite{Maynard1976}, and the fit function (black line) given by Eq.~\ref{Eq:rhos_rho}. The inset of (b) is a frequency sweep across the resonance, which shows the real part (black circles) and imaginary part (blue circles) of the relative dielectric constant measured at 2 bar and 1.62 K and the corresponding fit functions obtained from Eq.~\ref{eq:harm_osc}.}
\label{Fig_4}
\end{figure}

The resonance frequency $\omega_0/2\pi$ and quality factor $Q$ of the superfluid oscillator, from 1.6 K to $T_{\lambda}$ at various pressures, are shown in Fig.~\ref{Fig_4}. Combined with Eq.~\ref{Eq:resonance} we can then extract the superfluid fraction, $\rho_s/\rho$, shown in Fig.~\ref{Fig_4}(c). Calculations~\cite{Fisher1972} and measurements~\cite{Goldner1992, Rhee1989}, suggest the following functional form for $\rho_s/\rho$:
\beq\label{Eq:rhos_rho}
\frac{\rho_s}{\rho} = k(1+D_\rho t^{\Delta})t^{\zeta}  
\eeq
with $k = k_0(1+k_1t)$, and $t=(T-T_{\lambda})/T_{\lambda}$ the reduced temperature.  $k_0$, $k_1$, and $D_\rho$ are pressure dependent fit parameters of the critical behavior, $\zeta=0.6705\pm0.0006$ the critical exponent of the superfluid fraction~\cite{Goldner1992} and $\Delta=0.5$ - details of the fit parameters can be found in Appendix F. We find good agreement with Eq.~\ref{Eq:rhos_rho} at all pressures, demonstrating the universality of the lambda transition in our data. An exciting implication is that by replacing \he4 by \he3 in this superfluid nanoresonator, one could measure the superfluid fraction $\rho_s/\rho$ of confined superfluid \he3, which, according to Wu~\etal~\cite{Wu2013}, will lead to a direct signature of the Majorana surface excitations.  

\begin{figure}[t]
\centering
\includegraphics[width=8.6cm]{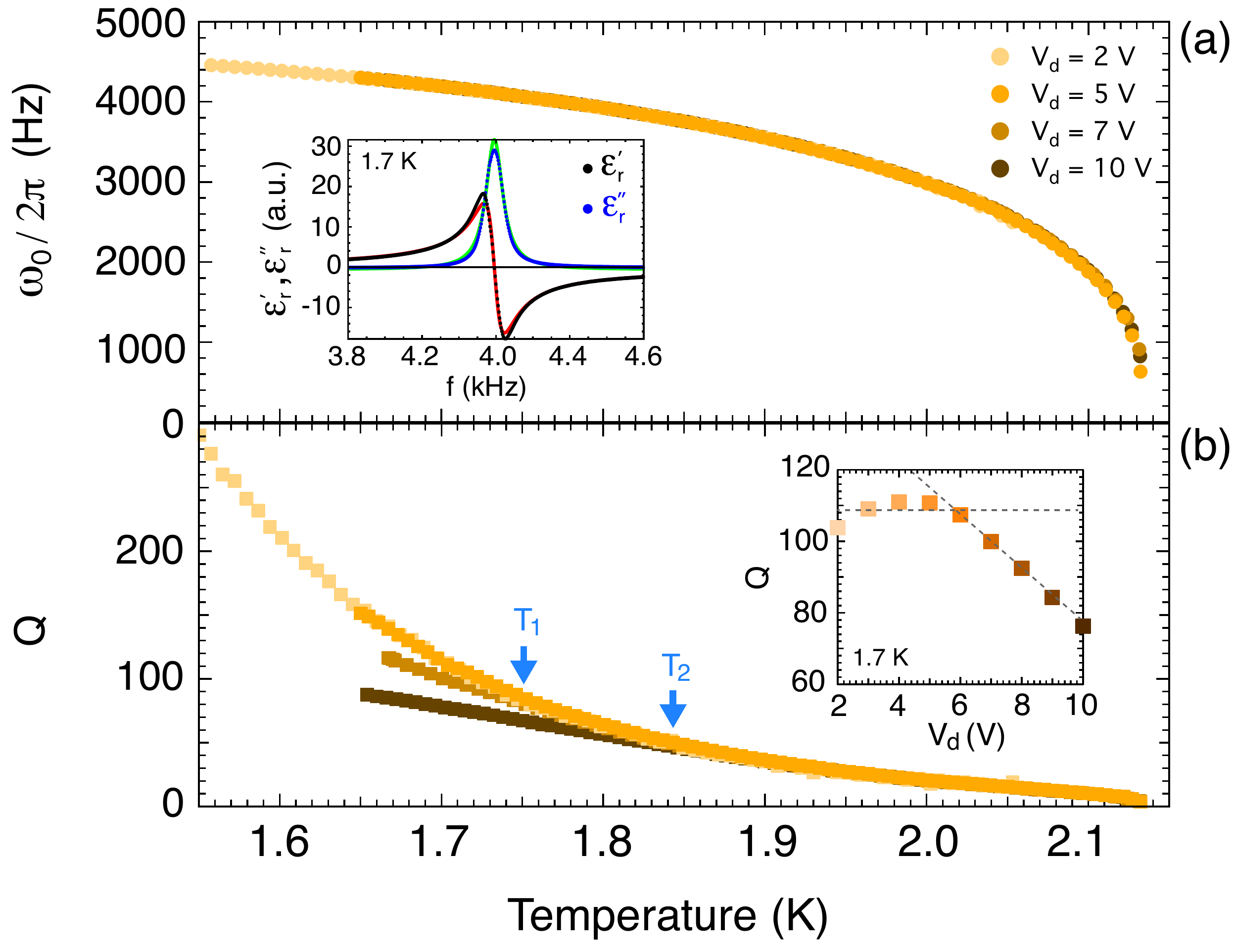}
\caption{Temperature dependence of the superfluid nanomechanical resonator  (a) frequency and (b) quality factor, taken at constant pressure $P=5$ bar for various drive amplitudes: 2, 5, 7 and 10 V. The inset of (b) shows the quality factor measured at $T=1.70$ K (black arrow) as a function of the drive voltage. The inset of (a) is a frequency sweep of the in-phase (black circles) and the out-of-phase (blue circles) dielectric response at $V_d=10$ V with their fit functions obtained from Eq.~\ref{eq:harm_osc}.}
\label{Fig_5}
\end{figure}

We note that the agreement between the data taken at a low drive voltage ($V_d\leq5$ V) and the fit shown in the inset of Fig.~\ref{Fig_4}b, is an indication of the linear behavior of the oscillator. We also measured the resonance at various drive voltages to explore deviations from the linear regime. We show in Fig.~\ref{Fig_5}b that at low drive ($V_d\le7$ V), the data for the quality factor as a function of temperature collapse on the same curve. At higher drive ($V_d\ge7$ V) they deviate from that curve at particular temperatures ($T_1\simeq1.75$ K for $V_d=7$ V and $T_2\simeq1.85$ K for $V_d=10$ V). This indicates a temperature dependent drive threshold. As can be seen in the inset of Fig.~\ref{Fig_5}b, the quality factor measured at $T=1.7$ K as a function of drive shows a threshold near $V_d\sim6$ V. Above this threshold, the dissipation increases because the flow in the channel enters a regime of quantum turbulence~\cite{Vinen2002}.

The resonance frequency data remain unchanged even at the highest drive, indicating that we are only in a slightly nonlinear regime and so we can still fit the resonance curve with Eq.~\ref{eq:harm_osc} to a good approximation (inset of Fig.~\ref{Fig_5}a). This allows us to calculate when the velocity of the oscillating mass in the channel passes above the critical velocity, $v_c$.  For instance, at $T=1.85$ K and $V_d=10$ V, the average superfluid velocity in the channel at resonance is given by
\beq
v_s=\frac{1}{\sqrt{2}}\frac{A\rho}{a\rho_s}\frac{F_{\rm stat}}{k_{\rm plate}}\omega_0 Q\sim 14\ \mathrm{m/s}
\eeq
with $Q\simeq50$, $F_{\rm stat}\simeq7\times10^{-3}$ N, and $\omega_0/2\pi\simeq4$ kHz. This is larger than what has been measured by Clow~\etal~\cite{Clow1967} in porous materials with a pore diameter of 200 nm, where they find a critical velocity of $\sim1$ m/s at 1.85 K.  
Exceeding the critical velocity results in the formation of quantum turbulence, which is known to decay through a cascade to smaller vortices - the Kolmogorov spectrum~\cite{Kolmogorov1991}.  Our superfluid nanoresonators may allow the study of quantum turbulence in a new regime, where vortices become pinned by the confined geometry and therefore change this vortex decay, as compared with bulk turbulence.  

Finally, we present additional experiments that could be performed using these superfluid nanomechanical resonators. One can identify two limits in the mechanical system presented above, depending on the ratio spring constants ($\Sigma$). In the ``soft plate" limit ($\Sigma\ll1$), the effective stiffness $k_{\rm plate}/(1+\Sigma)$ reduces to the stiffness of the plate $k_{\rm plate}$ only, and does not depend on thermodynamic variables $(T,P)$ of the liquid.  That is, one could remove the compressibility of the superfluid from Eq.~1, which help to isolate the  temperature and pressure dependence of the superfluid fraction. 

In the other ``stiff plate" ($\Sigma\gg1$) limit, the resonant frequency reduces to the formula for a fourth-sound Helmholtz resonator~\cite{Kriss1968}, $\omega_h^2=c_4^2{a}/(l_{\rm eff}V_{\rm cav})$, where $c_4=\sqrt{\rho_s/(\rho^2\chi)}$ is the fourth-sound velocity of liquid \he4~\cite{Revzen1974,Tam1985}, a sound mode that propagates only in the superfluid phase when the normal component is clamped. One can possibly drive this mode in our resonator using an electrostrictive driving force. The work presented here is between these two limits ($\Sigma\sim0.1$); details given in appendix G.

\section{\label{sec:level1}Conclusion}
We have presented devices to explore quantum fluids under nanoscale confinement - probing just nanoliters of superfluid with a high signal-to-noise. In the low-frequency limit, we measured the dielectric constant, and therefore the total density, of liquid \he4, which set the stage for probing the resonant behavior of the confined superfluid. In the dynamic regime, the device is a superfluid Helmholtz resonator, with a scale of tens to hundreds of nanograms of oscillating liquid \he4.  We used an analytical model to describe its dynamics, and we performed experiments to measure the superfluid fraction and the onset of quantum turbulence.  This system provides opportunities to study superfluids in restricted geometries - such as measuring the superfluid fraction in \he3, which will provide a direct signature of Majorana fermions at the surfaces - as well as providing opportunities for studying nanomechanical resonators at low temperatures with intrinsic quantum properties.  

\section*{\label{sec:level1}Acknowledgements}
This work was supported by the University of Alberta, Faculty of Science; the Natural Sciences and Engineering Research Council, Canada; the Canada Foundation for Innovation; Alberta Innovates Technology Futures; and the Alfred P. Sloan Foundation.  We thank G.G. Popowich for technical assistance, J.R. Beamish for his help regarding the capacitance measurement technique and a critical reading of the manuscript, and J. Maciejko for helpful discussions. We are particularly indebted to K.C. Schwab for numerous insightful conversations and originally pointing out the existence of a Helmholtz resonance in our system.

\appendix
\section{Nanofabrication}

An important component of our experiment is the realization of very well defined nanofluidic structures using cleanroom techniques. Here, the design of the nanofluidic device is a cylindrical basin (radius $r_{\rm cav}=2.5$ mm and height $h_{\rm cav}=1100$ nm) and four channels (length $l_{\rm cha}=2.5$ mm, width $w_{\rm cha}=1.6$ mm and height $h_{\rm cha}=550$ nm). Electrodes (height $h_{\rm ele}=100$ nm and radius $r_{\rm ele}=2$ mm) were deposited on both inner sides of the cavity to form a nanofluidic capacitor. In this section, we describe in detail the nanofabrication process of our devices.

The process starts, Fig.~\ref{Fig_1}(a), with the deposition of a Cr/Au masking layer (30 nm/180 nm) on a 100 mm x 100 mm x 1.1 mm borosilicate glass wafer previously cleaned with a piranha solution (3:1 H$_2$SO$_4$ and H$_2$O$_2$). A first optical lithography is performed to pattern the design of the cavity and the channels. For that, a positive photoresist polymer (HPR504) is spun onto the wafer (10 s at 500 RPM and then 40 s at 4000 RPM) and baked for 30 min at 115 degrees, leading to a thickness of 1.2 $\mu$m. The photoresist is exposed for 2.2 s with UV light (365 nm) through the photomask and developed, and as a result the photomask pattern is transferred onto the photoresist. We chemically etch the exposed masking layer (Cr/Au) with an acidic solution and then the glass wafer down to a certain depth (550 nm for this device) with a glass etchant (50\% HF, 10\% nitric acid and 40\% water).

At this point, the nanofluidic cavity and channels are etched in the glass wafer. Afterwards, the photoresist and masking layers are stripped off and a second optical lithography is performed in order to pattern the electrodes. For that, we repeat the steps described above with a second photomask. Next, using a sputtering system we deposited a Cr/Au thin film (10 nm/90 nm) on the wafer. We then lift off the photoresist to obtain the electrode pattern in the bottom of the cavity and channels, Fig.~\ref{Fig_1}(b).

At this stage, we dice the wafer into smaller rectangular pieces (10 x 15 mm). These pieces are piranha cleaned and bonded using direct bonding, which consist of an additional soft mechanical cleaning of the pieces with a soap solution and the pressing by hand under the microscope of the two pieces against each other. This finishes the nanofabrication, and the relevant confinement length of this device, $h_{\rm eff}$, is given by the distance between the electrodes, Fig.~\ref{Fig_1}(c).

We bond the two rectangular pieces perpendicularly, Fig.~\ref{Fig_1}d, such that we can solder electrical wires to the electrodes that are deposited in the bottom of the cavity. The device is then placed in a copper sample cell and connected to the electrical coaxial feedthroughs. Next, this sample cell is sealed with an indium o-ring and mounted on a cryostat.

\begin{figure}[b]
\centering
\includegraphics[width=8.0cm]{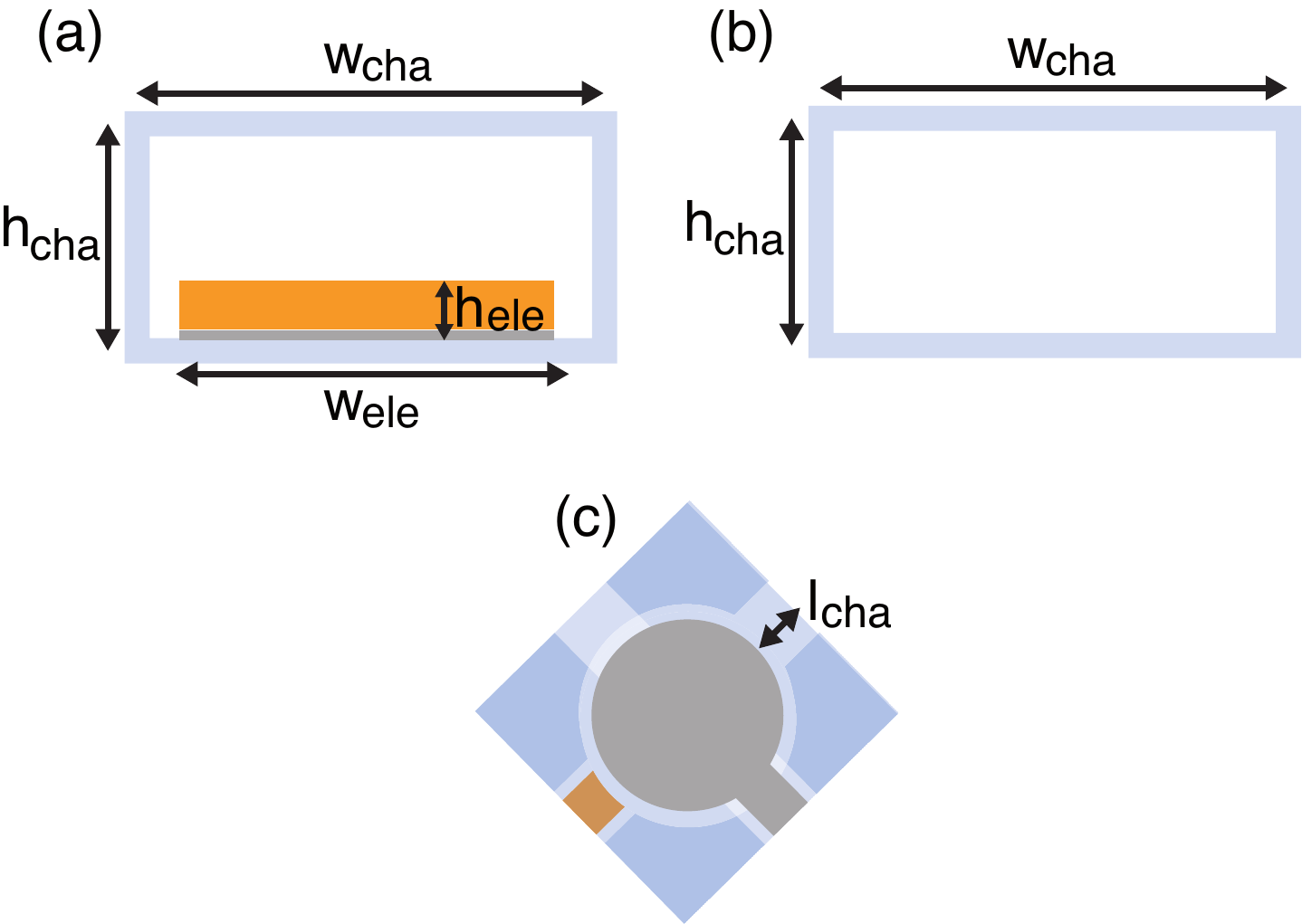}
\caption{Cross section of the channels having an electrode passing through (a) and without an electrode (b). The region of the nanofluidic device confining the liquid \he4: a cavity and four channels (c). }
\label{Fig_S2}
\end{figure}

We study the properties of the flow in the four nanofluidic channels defined by this nanofabrication process. The four channels have the same dimensions but because two channels have electrodes passing through they have different cross sectional geometry. We show in Fig.~\ref{Fig_S2} the cross sections of the two type of channels.

\section{Characterization}
During the nanofabrication process of our devices and prior to the bonding, we control the depth of glass etch and the thickness of metal depositions with a surface profiler (Alpha Step IQ). It is a diamond tip on a piezoelectric transducer, which is brought in contact with the device and moved laterally across it to measure the topography. This tool has an excellent resolution ($\sim1$ nm).

We have also used a second technique based on optical interference to characterize the confinement length after bonding. This technique has been described previously~\cite{Rojas2014} and allowed us to precisely measure the uniformity of the confinement length ($\sim1$ \%).

\section{Control of temperature and pressure}
We filled the sample cell and the nanofluidic structure with liquid \he4 of natural purity ($\sim300$ ppb \he3). To regulate the pressure in the cell, the fill line is filled with a gas pressure and connected to a ballast containing a heater, which is dipped into liquid nitrogen. A pressure gauge (Mensor CPT 6000) with a precision of 2 mbar is connected to the fill line, and a proportional-integral-derivative (PID) controller allows us to regulate the temperature of the gas in the ballast in order to maintain the pressure in the fill line.

We measure the temperature of the sample cell with a carbon glass resistive thermometer and a resistance bridge (LakeShore 370 AC). We regulate the temperature with a heater on the sample and the PID controller of the resistance bridge.

\section{Measurement of plate stiffness: $k_{\rm plate}$}
The stiffness of the glass plates $k_{\rm plate}$ can be computed from classic theory of plate elasticity. To do that precisely, one has to know all the mechanical properties at low temperature. Another possibility is to measure it directly, and this is what we have chosen to  do. We performed a measurement of the spacing between the plates via the capacitance while applying an electrostatic force between the electrodes. We applied a varying dc voltage ($V_{\rm dc}=0 - 20$ V) across the electrodes on top of an ac voltage ($V_{\rm ac}=1$ V), so the drive voltage is
\beq
V_{d}(t)=V_{\rm dc}+V_{\rm ac}\cos{\omega t}
\eeq
and the electrostatic force between the electrodes
\beq\label{eq.Fstat1}
F_{\rm stat}(t)=\frac{1}{2}\epsilon_0\epsilon_r \frac{A_{\rm ele}}{h_{\rm eff}^2}V_d(t)^2.
\eeq
This force leads to a deflection of the glass plates inversely proportional to their bending stiffness. One can write the relation between the average deflection across the electrodes $x(t)$ and the electrostatic force,
\beq\label{eq.Fstat2}
F_{\rm stat}(t)=k^{\prime}_{\rm plate}x(t)
\eeq
with $k^{\prime}_{\rm plate}=k_{\rm plate}/(1+\beta)$ and $\beta$ a factor related to the radius difference between the electrodes and the glass plates, $\delta r=r_{\rm cav}-r_{\rm ele}=500\ \mu$m. By integrating the standard expression for the bending of a circular plate~\cite{Blaauwendraad2010} over the plate surface area, we find
\beqn
\beta & =  & 2\frac{\delta r}{r_{\rm cav}}+3\left(\frac{\delta r}{r_{\rm cav}}\right)^2-4\left(\frac{\delta r}{r_{\rm cav}}\right)^3+\left(\frac{\delta r}{r_{\rm cav}}\right)^4\\
 & \simeq & 0.49.
\eeqn

\begin{figure}[h]
\centering
\includegraphics[width=8.6cm]{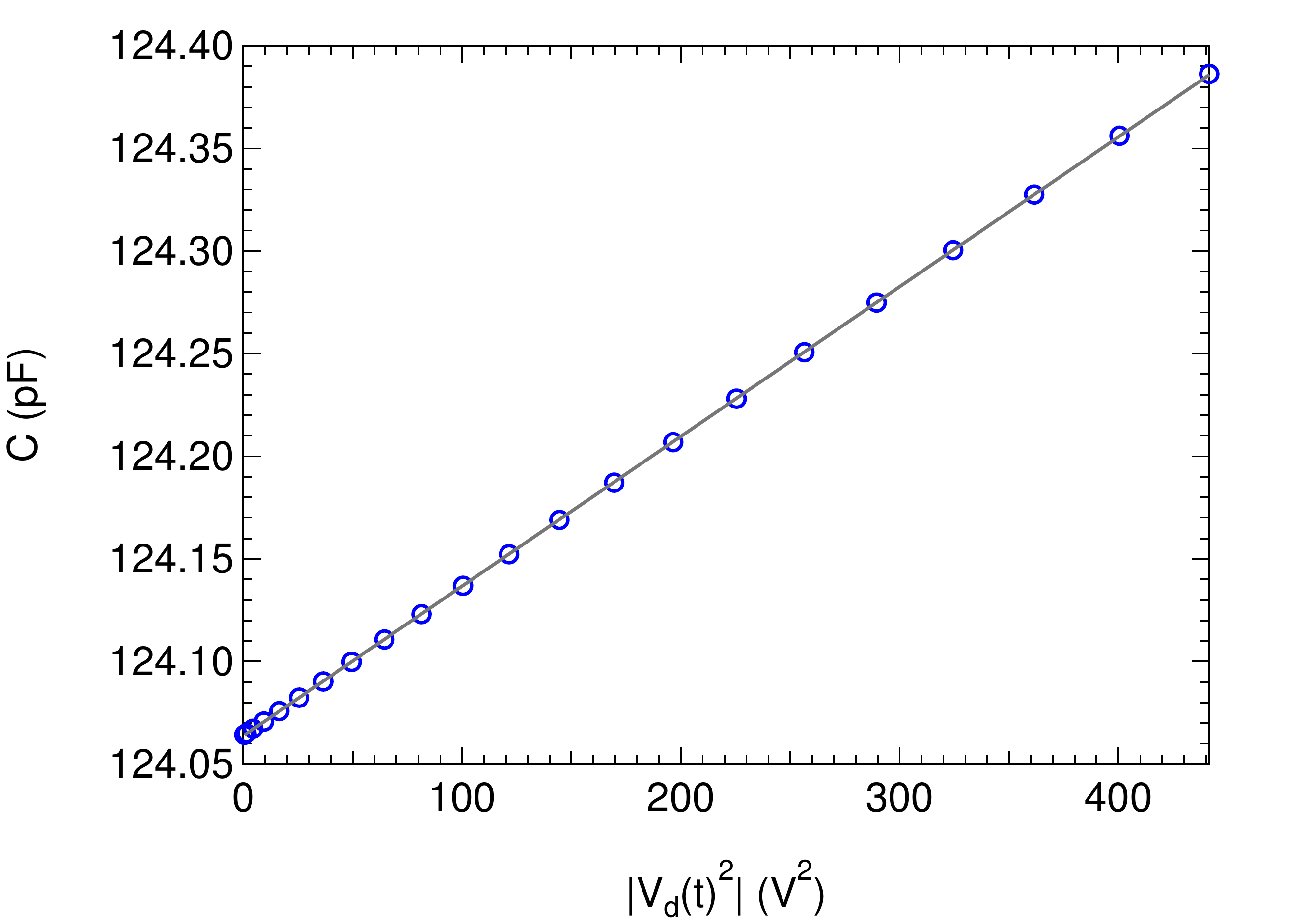}
\caption{Capacitance measurement (blue circles) of the nanofluidic capacitor under the application of a varying dc voltage (0-20 V), which bends the cavity glass plates and increases the capacitance $C$. These data are fit (gray line) to Eq.~\ref{eq.capafit} in order to extract the bending stiffness of the glass plates.}
\label{Fig_S3}
\end{figure}

The deflection of the glass plates is related to a change in capacitance. To the first order in $x(t)$, the capacitance is given by
\beq\label{eq.capacitance}
C(t)=\frac{A\epsilon_0\epsilon_r}{h_{\rm eff}+x(t)}\simeq C_0+C_0\frac{x(t)}{h_{\rm eff}}.
\eeq
Substituting Eq.~\ref{eq.Fstat1} in Eq.~\ref{eq.Fstat2}, and Eq.~\ref{eq.Fstat2} in Eq.~\ref{eq.capacitance}, we obtain
\beq\label{eq.capafit}
C(t) = C_0+\gamma V_d(t)^2
\eeq
with
\beq
\gamma=\frac{1}{2}\frac{C_0^2}{h_{\rm eff}^2 k^{\prime}_{\rm plate}}.
\eeq
We show in Fig.~\ref{Fig_S3} a measurement of the capacitance as a function of the average applied voltage squared
\beq
\overline{V_{d}(t)^2}=V_{\rm dc}^2+\frac{V_{\rm ac}^2}{2},
\eeq
which allows us to extract $\gamma$ and therefore $k^{\prime}_{\rm plate}$. In the measurement, Fig.~\ref{Fig_S3}, performed at $T=1.68$ K and with a cell filled with liquid \he4 at $P=5$ bar, we fit the data with Eq.~\ref{eq.capafit} and obtained $C_0=124.06$ pF and $\gamma=7.3\times10^{-4}$ pF/V$^2$, which leads to
\beq
k^{\prime}_{\rm plate}=\frac{1}{\gamma}\frac{1}{2}\frac{C_0^2}{h_{\rm eff}^2}\simeq1.30\times10^7\ \mathrm{N/m}
\eeq
and finally the bending stiffness of the glass plate is given by
\beq
k_{\rm plate}=(1+\beta){k^{\prime}_{\rm plate}}\simeq1.94\times10^7\ \mathrm{N/m}.
\eeq

\section{Equations of motions for the superfluid nanomechanical resonator}

Our nanofluidic structure is composed of a cylindrical cavity connected to four channels of a rectangular cross section. The liquid \he4 that filled the nanofluidic structure has natural acoustic resonances. In addition, the cavity walls are flexible and have drum-like resonant modes. To completely describe this superfluid nanomechanical resonator, one has to take into account the acoustic resonances of the liquid \he4, the mechanical resonances of the nanofluidic structure itself, and the coupling between these modes. However, we can make useful approximations and obtain a simple analytical model that describes satisfactorily the superfluid resonance.

We first assume that the cavity walls are rigid. In this case, there is a resonance related to the oscillation of the superfluid in the channels and the compression of the fluid in the cavity. This Helmholtz resonance is analogous to a mass-on-a-spring, with the potential energy stored by the fluid in the cavity and the kinetic energy stored by the fluid oscillation in the channels. This description is valid if the dimensions of the structure are smaller than the acoustic wavelength $\lambda_a$ in the fluid. In our experiment, the highest resonance frequency is $\omega_0/2\pi=5$ kHz and the smallest first sound velocity $c_1=230$ m/s, so the acoustic wavelength is $\lambda_a>46$ mm. Since the largest dimension in our geometry ($\sim 10$ mm), is about five times smaller than the smallest acoustic wavelength, $\lambda_a=46$ mm, this description is valid. In our experiment, the cavity walls are flexible and have drum-like resonances at much higher frequency ($\sim 100$ kHz) than the Helmholtz resonance of the fluid ($\omega_0/2\pi<5$ kHz). As a result, these modes do not hybridize significantly and, near the Helmholtz resonance of the fluid, we can reasonably assume that the effect of the flexible cavity walls is only to redefine the stiffness constant $k_{\rm plate}$.
\begin{figure}[h]
\centering
\includegraphics[width=7cm]{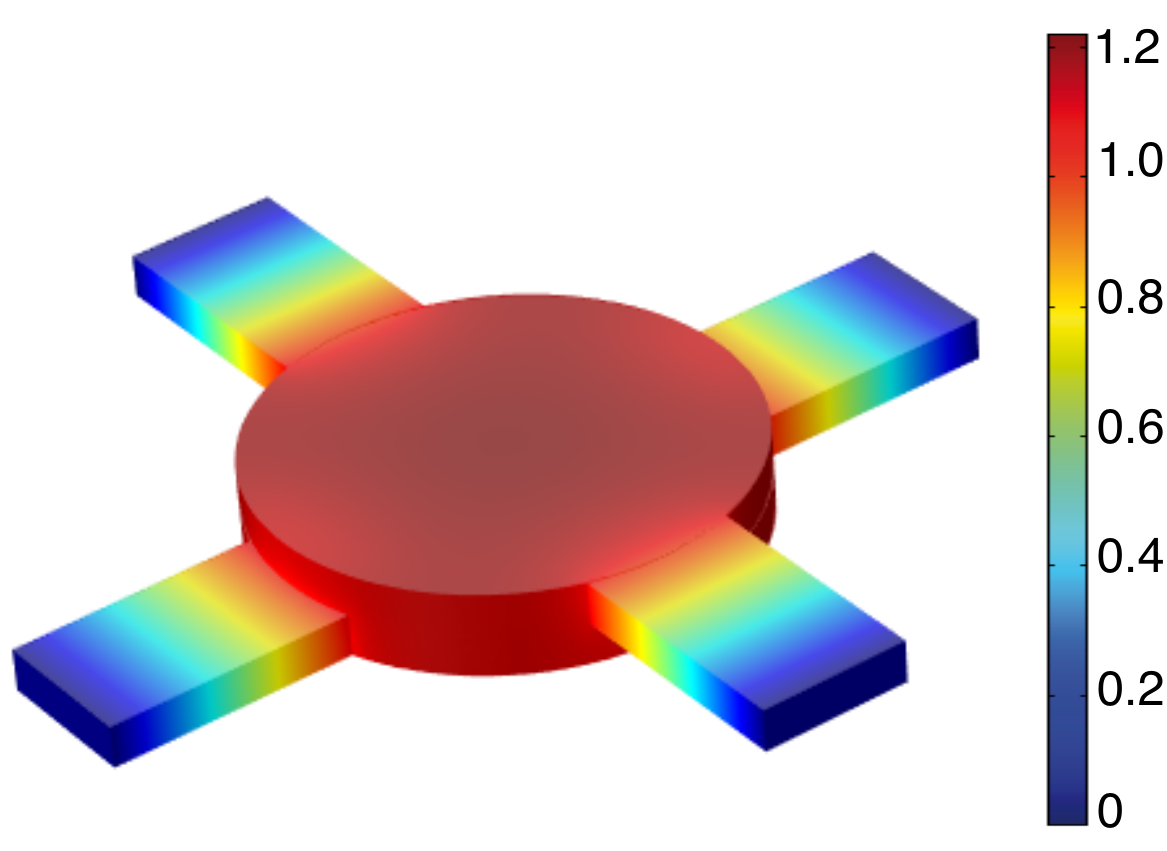}
\caption{FEM simulation of the first mode of liquid \he4 confined in the nanofluidic device. The color bar represents the acoustic pressure on resonance ($\omega_0/2\pi\sim7$ kHz) in arbitrary units. The mode shape is analogous to the Helmholtz resonance with four masses in the channels connected to an effective spring in the cavity. In this simulation, we increased the thickness of the structure by a factor of 1000 so the mode shape is easier to see.}
\label{Fig_S4}
\end{figure}

We show in Fig.~\ref{Fig_S4} a finite-element-method (FEM) simulation of the acoustic mode of the liquid \he4 confined in the nanofluidic device. The mode shape is similar to the Helmholtz resonance with four masses in the channels connected to an effective spring in the cavity. To find the resonance frequency of this mode, we write the kinetic and potential energy for this mechanical system. 

All the kinetic energy is concentrated in the vicinity of the channel flow where the velocity is the largest. Since the channel confinement length is smaller than the viscous penetration depth ($h_{\rm cha}<\lambda_\nu$), the normal component is clamped and only the superfluid can oscillate. In our nanofluidic structure, there are two pairs of channels of the same dimensions; one type of channel has an electrode passing through, and the other one does not. This leads to the following kinetic energy:
\beq
E_K=\frac{1}{2} \rho_s l (a_1 v_1^2 + a_2 v_2^2)2
\eeq
where $a_1=8.2\times10^{-10}$ m$^2$ and $a_2=8.8\times10^{-10}$ m$^2$ are the cross sectional areas of the channel type with an electrode and without an electrode, respectively. $l=2.5\times10^{-3}$ m is the effective length of the channels, a sum of the physical length of the channel plus a correction due to effects of the diverging flow at the ends of the channel. This correction factor scales like the cross sectional area of the channel, which is small in this case, and so this correction factor will be neglected in our analysis. $v_1$ and $v_2$ are the average velocities of the superfluid in the two different types of channels, which are related by $a_1v_1=a_2v_2$ from conservation laws. As a result, we can define an effective superfluid velocity $v$ such that $v=v_1=(a_2/a_1)v_2$. Hence, we have
\beq\label{eq.Ek}
E_K=\left(1+\frac{a_1}{a_2}\right) \rho_s l a_1 v^2.
\eeq
This expression represents the kinetic energy of an effective mass,
\beq\label{eq.meff}
m_{\rm He}= 2 \left(1+\frac{a_1}{a_2}\right) \rho_s l a_1 
\eeq
moving at a velocity $v$.

The potential energy is stored in the deflection of the glass plates and the compressibility of the liquid confined in the cavity. The compressibility of the liquid outside the nanofluidic device is much larger due to the volume difference, and so it does not contribute. The potential energy can be written
\beq\label{eq.Ep}
E_P=\frac{1}{2} \frac{k_{\rm plate}}{1+\Sigma} x^2
\eeq
with $\Sigma=k_{\rm plate}/k_{\rm He}$, $k_{\rm He}=A_{\rm plate}^2/(\chi V_{\rm cav})$, $\chi$ the compressibility of liquid \he4, $A_{\rm plate}$ the surface area of the cavity glass plates, $V_{\rm cav}$ the volume of the cavity, $k_{\rm plate}$ the stiffness of the glass plates and $x$ the change in the cavity height induced by the deflection of the glass plates. The conservation of mass leads to the following relation
\beq
xA\rho= 2 \rho_s a_1 \left(1+ \frac{a_1}{a_2}\right) y,
\eeq
where $y$ is the effective displacement of the superfluid mass in the channels. Hence, the potential energy can be written
\beq
E_P= 2\frac{k_{\rm plate}}{1+\Sigma} \left(\frac{\rho_s a_1}{\rho A_{\rm plate}}\right)^2\left(1+ \frac{a_1}{a_2}\right)^2 y^2,
\eeq
which represents the potential energy of an effective spring, 
\beq\label{eq.keff}
k_{\rm eff}=4\frac{k_{\rm plate}}{1+\Sigma} \left(\frac{\rho_s a_1}{\rho A_{\rm plate}}\right)^2\left(1+ \frac{a_1}{a_2}\right)^2
\eeq
with an elongation $y$. The dynamics of the system is then simply described by the superfluid mass $m_{\rm He}$ attached to an effective spring $k_{\rm eff}$. Using Eq.~\ref{eq.meff} and Eq.~\ref{eq.keff} the resonance frequency of this mechanical system becomes
\beq
\omega_0^2=\frac{k_{\rm eff}}{m_{\rm He}}=2\left(1+ \frac{a_1}{a_2}\right)\left(\frac{\rho_s}{\rho}\right)\frac{a_1}{A_{\rm plate}^2 \rho l}\frac{k_{\rm plate}}{1+\Sigma}.
\eeq
In addition, since $a_1\simeq a_2$, we have
\beq\label{eq.omega0}
\omega_0^2\simeq4\left(\frac{\rho_s}{\rho}\right)\frac{a_1}{A_{\rm plate}^2 \rho l}\frac{k_{\rm plate}}{1+\Sigma}.
\eeq
Using Eq.~\ref{eq.omega0} and the bulk thermodynamic data of the density $\rho(T,P)$ and compressibility $\chi(T,P)$ obtained by Maynard~\cite{Maynard1976}, we extract $\rho_s/\rho$ from the resonance frequency measurements $\omega_0/2\pi$ of Fig.~4. We compared this data with Maynard's values for $\rho_s/\rho$ and found a good agreement if we add a correction factor $\alpha=0.42$, such that
\beq\label{eq.omega02}
\frac{\rho_s}{\rho}=\alpha\frac{\omega_0^2(1+\Sigma)}{2k_{\rm plate}}\frac{A_{\rm plate}^2 \rho l}{a_1\left(1+ \frac{a_1}{a_2}\right)}.
\eeq
Since this correction factor is the same at every pressure between 2 and 25 bar and every temperature between $T_\lambda$ and 1.6 K, it is related to the over simplified analytical model used here, which, for example, does not take into account the exact mode shape of the superfluid resonance.

\section{Superfluid Fraction in liquid \he4}
\begin{figure}[t]
\centering
\includegraphics[width=7cm]{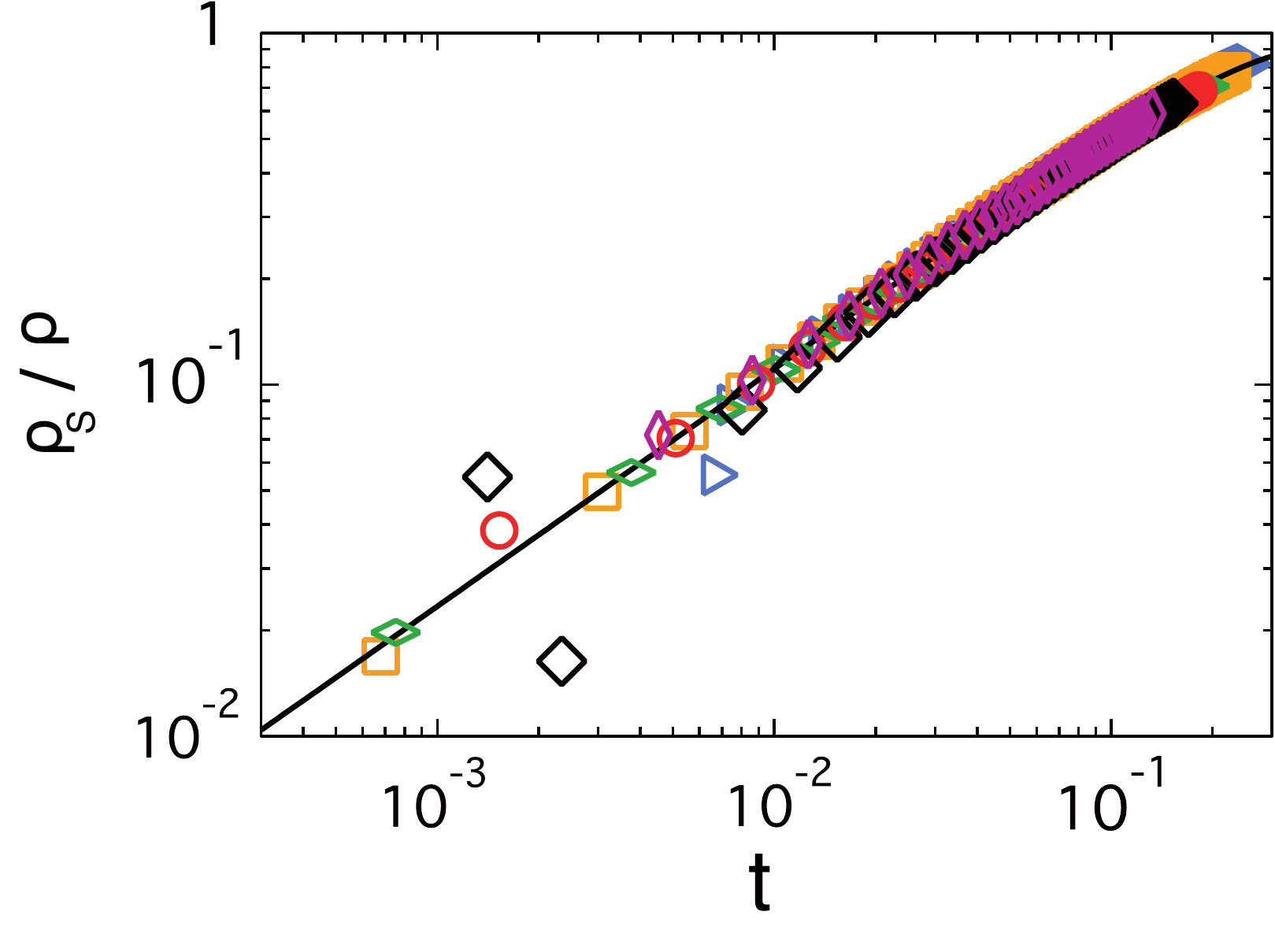}
\caption{Temperature dependence of the superfluid fraction, $\rho_s/\rho$, extracted from the resonance frequency using Eq.~\ref{Eq:resonance} with $T_\lambda$ taken from Maynard sound measurements~\cite{Maynard1976}. Data taken at a drive voltage $V_d=5$ V, for various pressures: 2 (blue triangles), 5 (orange squares), 10 (green diamonds), 15 (red circles), 20 (black diamonds), and 25 (purple diamonds) bar. The black line is obtained by fitting the data measured at 2 bar with Eq.~\ref{Eq:rhos_rho_S}.}
\label{Fig_S5}
\end{figure}

The study of thermodynamic functions (specific heat, superfluid fraction, compressibility, etc.)~at the superfluid transition of \he4 has provided an important test for the theory of critical phenomena~\cite{Singsaas1984,Lipa2003}. The bulk behavior of the superfluid fraction $\rho_s/\rho$ is well known~\cite{Goldner1992}, but very close to $T_\lambda$, finite-size effects can be revealed with nanoscale confinement~\cite{Rhee1989} and these effects are still not fully understood~\cite{Gasparini2008}. We show (Fig.~\ref{Fig_S5}) our measurements of the superfluid fraction as a function of the reduced temperature. In the bulk regime, previous works suggest the following functional form
\beqn\label{Eq:rhos_rho_S}
\frac{\rho_s}{\rho} & = & k(1+D_\rho t^{\Delta})t^{\zeta}\\
k & = & k_0(1+k_1t)\notag
\eeqn
with $t=(T-T_{\lambda})/T_{\lambda}$ the reduced temperature, $k_0$, $k_1$, and $D_\rho$ the pressure dependent fit parameters of the critical behavior, $\zeta$ the critical exponent of the superfluid density fraction, and $\Delta=0.5$ a fixed parameter (for a discussion about $\Delta$ see Ref.~\citenum{Singsaas1984}). This functional form has been used by Goldner~\etal\cite{Goldner1992} for the reduced temperature range $3\times10^{-7}<t<10^{-2}$, and they obtained the fit parameters given in Table~\ref{tab:tablefit}. The critical exponent $\zeta$ is universal and does not depend on the details of the experiment (\textit{i.e.}, liquid pressure), so for our fit analysis, we fixed $\zeta$ to the values obtained by Goldner~\etal\cite{Goldner1992} which are $\zeta=0.6705\pm0.0006$. In our case, the reduced temperature range is $2\times10^{-3}<t<2.5\times10^{-1}$, and using their best fit parameters values for $k_0$, $k_1$, and $D_\rho$, the functional form starts to deviate from our data near $t\simeq5\times10^{-2}$. In order to find a better agreement with the functional form of Eq.~\ref{Eq:rhos_rho_S}, we left the parameters $k_0$, $k_1$, and $D_\rho$ as fit parameters. The best-fit parameters obtained for various pressures are shown in Table~\ref{tab:tablefit}.

\begin{table}[t]
\caption{\label{tab:tablefit} The best-fit parameters using the functional form defined in Eq.~\ref{Eq:rhos_rho_S}. for fitting the superfluid fraction data.}
\begin{ruledtabular}
\begin{tabular}{ccccc}
 Pressure & $k_0$ & $k_1$ & $D_\rho$ & $\zeta$\\
\colrule
SVP~\footnote{Goldner~\etal~\cite{Goldner1992} at saturated vapor pressure (SVP)} & $2.38$ & $-1.74$ & $0.396$ & $0.6705$\\
$P=2$ bar~\footnote{This work} & 2.38 & $-1.06$ & 0.347 &$0.6705$\\
$P=5$ bar~\footnotemark[2] & 2.24 & $-1.17$ & 0.568 & $0.6705$\\
$P=10$ bar~\footnotemark[2] & 2.14 & $-1.30$ & 0.757 & $0.6705$\\
$P=15$ bar~\footnotemark[2] & 2.14 & $-1.36$ & 0.774 & $0.6705$\\
$P=20$ bar~\footnotemark[2] & 1.99 & $-1.62$ & 1.315 & $0.6705$\\
$P=25$ bar~\footnotemark[2] & 2.14 & $-1.55$ & 0.930 & $0.6705$
\end{tabular}
\end{ruledtabular}
\end{table}

We obtained relatively good agreement with the values obtained by Goldner~\etal~\cite{Goldner1992}, especially for the data at low pressure ($P=2$ bar), which is closer to the saturated vapor pressure near $T_\lambda$ ($P\sim0.05$ bar) used in Goldner's experiment.
Future analysis at lower reduced temperature $t$ may allow a detection of finite-size effects in the superfluid fraction. These effects, with the confinement length of our channels ($h_{\rm cha}\sim500$ nm), should appear near $t\sim10^{-4}$.

\section{Summary of Forces}
The superfluid nanomechanical resonator can be driven by various forces. As we described above, the voltage applied between the electrodes generates an electrostatic force between the glass plates, which is given by
\beq
F_{\rm stat}=\frac{1}{2}\epsilon_0\epsilon_r A_{\rm ele} E_d^2
\eeq
with $E_d=V_d/h_{\rm eff}$ the electric field between the electrodes, $\epsilon_0$ the vacuum permittivity and $\epsilon_r$ the dielectric constant of liquid \he4. In addition, this electrostatic field generates a pressure gradient in the liquid given by
\beq\label{eq.EHeffects}
\mathbf{\nabla}P=-\frac{\epsilon_0 E^2}{2}\mathbf{\nabla}\epsilon_r+\frac{\epsilon_0}{6}\mathbf{\nabla}\left[E^2\rho\frac{\mathrm{d}\epsilon}{\mathrm{d}\rho}\right].
\eeq
This electrohydrodynamic effect has been previously described for the general case~\cite{Stratton1941} and for the case of superfluid \he4~\cite{Jackson1982}. In our case, we can reasonably assume that the dielectric constant is homogeneous between the electrodes, so the first term in Eq.~\ref{eq.EHeffects} can be neglected and only the second term (electrostriction) remains. In addition, since the Clausius-Mossotti relation can be used for a non-polar liquid such as \he4~\cite{Chan1977}, the second term of Eq.~\ref{eq.EHeffects} can be simplified. Hence, the pressure difference between the region outside the electrodes ($E_d=0$) and the region between the electrode ($E_d\neq0$) is given by
\beq
\Delta P_{\rm strict}=\frac{\epsilon_0}{6}(\epsilon_r-1)(\epsilon_r+2)E_d^2.
\eeq
As the electric field is increased, this pressure difference induces a flow from the channels towards the cavity. On the other hand, the electrostatic force induces a deflection of the cavity walls, which generates an increase of pressure in the cavity and a flow from the cavity toward the channels. These two effects are competing and, depending on the geometry of the nanofluidic structure, one can make an electrostatically or electrostrictively driven resonator.

\begin{figure}[t]
\centering
\includegraphics[width=8.6cm]{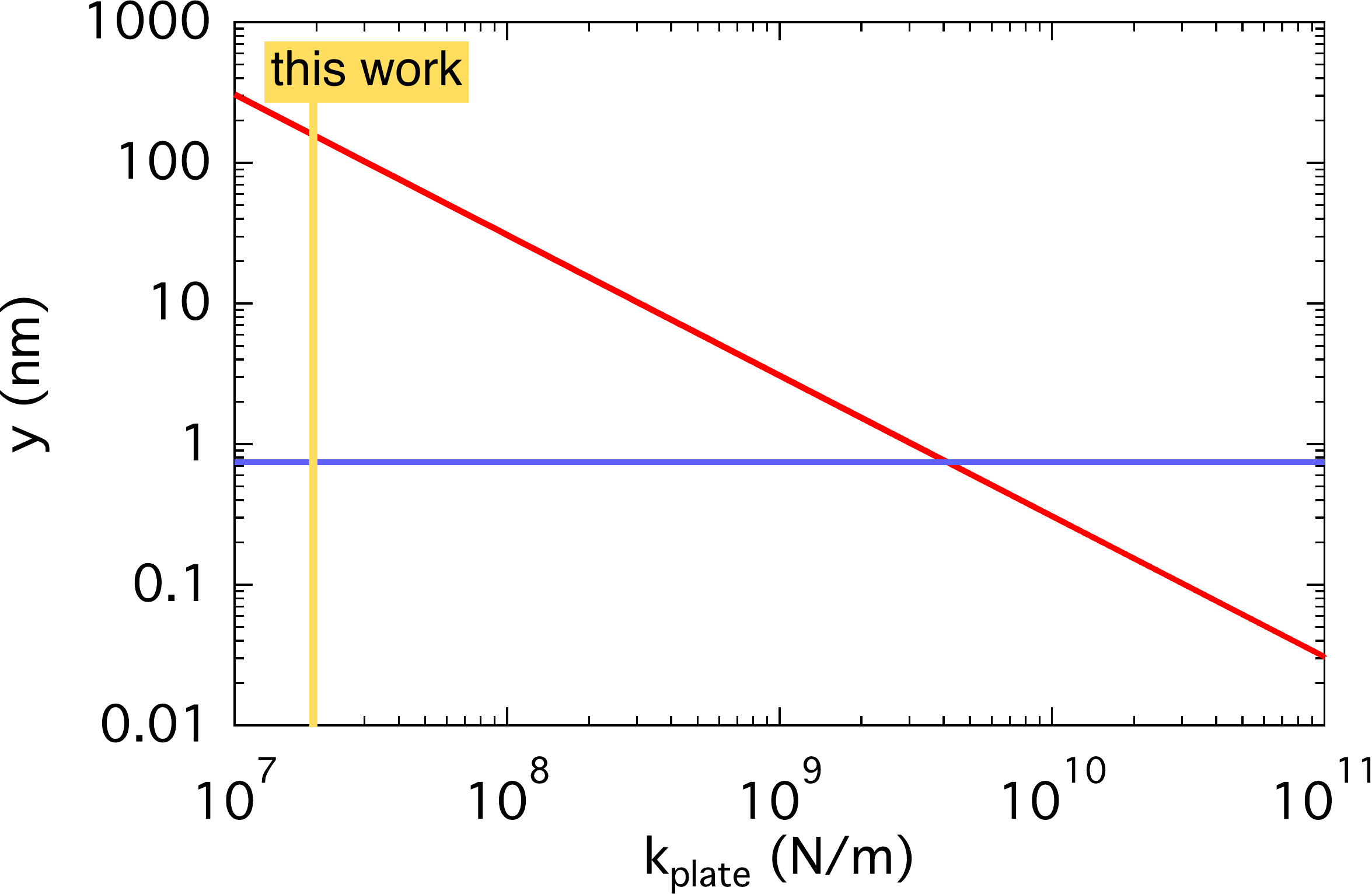}
\caption{Displacement of the superfluid mass $y$ induced by the electrostatic drive (red line) and the electrostrictive drive (blue line) as a function of the stiffness of the glass plates. In our geometry, the superfluid nanoresonator is mainly electrostatically driven.}
\label{Fig_S6}
\end{figure}

We now compare these two driving terms for our geometry. Off resonance, the electrostatic force induces a change in the cavity height given by
\beq
x=\frac{F_{\rm stat}}{k_{\rm plate}}
\eeq
so the displacement of the effective superfluid mass in the channels is
\beq
y_{\rm stat}=\frac{F_{\rm stat}}{k_{\rm plate}}\frac{\rho}{\rho_s}\frac{A_{\rm plate}}{2 a_1 \left(1+ \frac{a_1}{a_2}\right)}.
\eeq
On the other hand, the electrostriction induces a displacement of the superfluid mass in the channels given by
\beq
y_{\rm strict}=-\chi V_{\rm cav}\Delta P_{\rm strict}\frac{\rho}{\rho_s}\frac{1}{2a_1\left(1+\frac{a_1}{a_2}\right)},
\eeq
which can be rewritten
\beq
y_{\rm strict}=-\frac{F_{\rm strict}}{k_{\rm He}}\frac{\rho}{\rho_s}\frac{A_{\rm plate}}{2a_1\left(1+\frac{a_1}{a_2}\right)},
\eeq
with $F_{\rm strict}=A_{\rm plate}\Delta P_{\rm strict}$. To find the dominant effect, one can write
\beq
\frac{y_{\rm strict}}{y_{\rm stat}}=\frac{F_{\rm strict}}{F_{\rm stat}}\frac{k_{\rm plate}}{k_{\rm He}}
\eeq
and when $\Sigma=k_{\rm plate}/k_{He}\sim0.1$, as for our geometry, we have
\beq
\frac{y_{\rm strict}}{y_{\rm stat}}\simeq \frac{A_{\rm plate}\frac{\epsilon_0}{6}(\epsilon_r-1)(\epsilon_r+2)E_d^2}{\frac{1}{2}\epsilon_0\epsilon_r A_{\rm ele} E_d^2}0.1.
\eeq
For our geometry, this leads to $y_{\rm stat}>100y_{\rm strict}$.  As a result, in our experiment the resonator is mainly driven by the electrostatic force applied on the cavity walls. However, one can possibly obtain an electrostrictively driven resonator by increasing the stiffness of the plates, which can be done by reducing the cavity radius or increasing the thickness of the cavity glass walls. We show (Fig.~\ref{Fig_S6}) the displacement induced by the electrostatic and electrostrictive drive as function of the stiffness of the glass plates.

\begin{figure}[t]
\centering
\includegraphics[width=8.6cm]{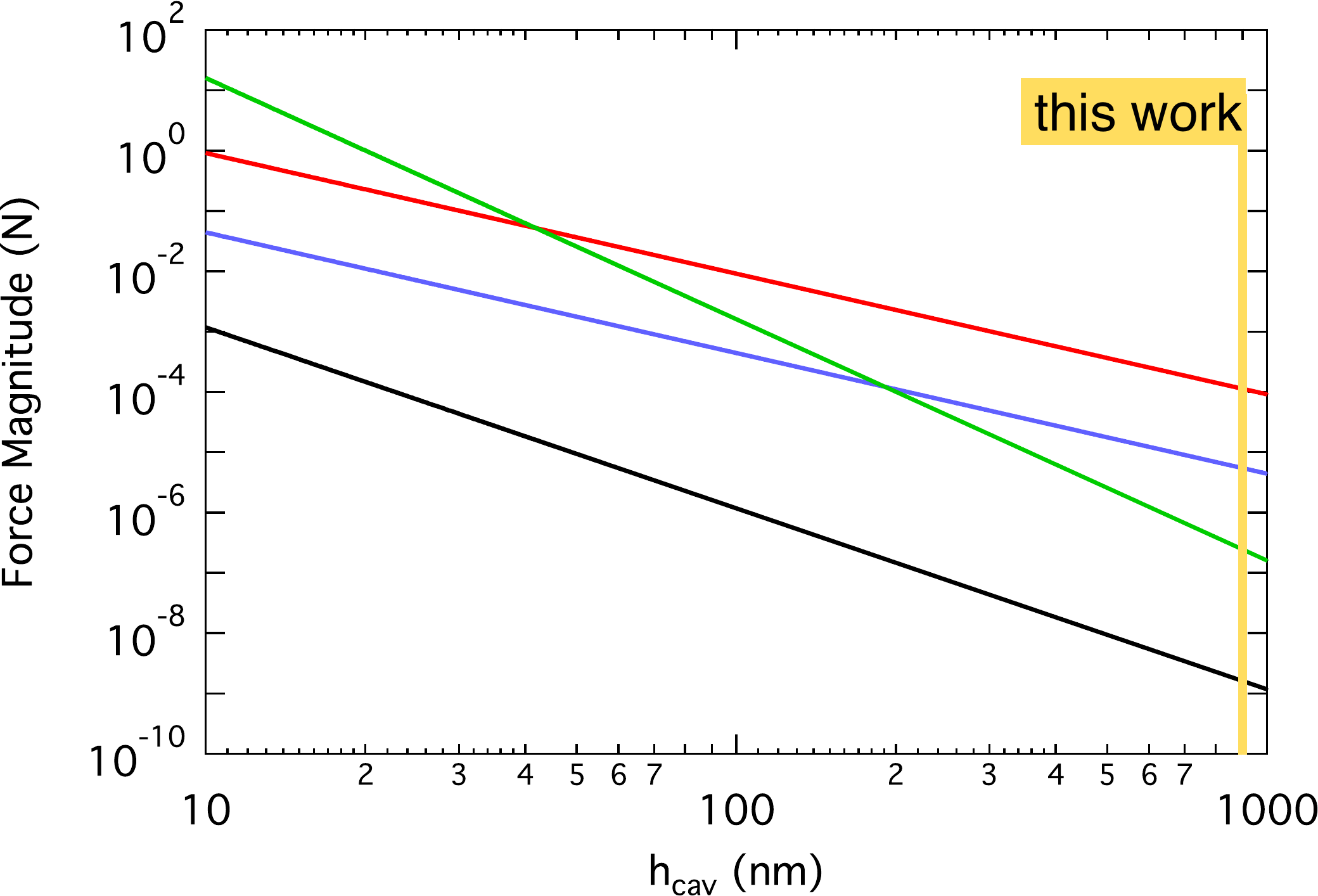}
\caption{Magnitude of the four forces acting on the superfluid nanomechanical resonator as a function of the gap $h_{\rm eff}$ between the electrodes for $V_d=1$ V. There is an electrostatic force $F_{\rm stat}$ (red line), an electrostrictive force $F_{\rm strict}$ (blue line), an electromagnetic Casimir force $F_{\rm Cas}^{\rm EM}$ (green line), and a critical Casimir force $F_{\rm Cas}^{\rm crit}$ (black line).}
\label{Fig_S7}
\end{figure}

Finally, other forces acting on the glass plate can drive the superfluid nanomechanical resonator. There is an attractive electromagnetic Casimir force~\cite{Casimir1948} between the electrodes, which is given by
\beq
F_{\rm Cas}^{\rm EM}=\frac{\pi^2}{240}\frac{\hbar c}{h_{\rm eff}^4}A_{\rm ele}
\eeq
with $\hbar$ Planck's constant and $c$ the speed of light. This force results from the confinement of the electromagnetic field vacuum fluctuations. Its magnitude is usually too small to be detected on the macroscopic scale, but it becomes non-negligible in nano/microstructures, as pointed out by Chan~\etal~\cite{Chan2001}. There is also a critical Casimir force given by
\beq\label{eq.critcas}
F_{\rm Cas}^{\rm crit}=\frac{2 k_{\rm B} T_\lambda}{h_{\rm eff}^3}A_{\rm plate},
\eeq
at $T=T_\lambda$. This force arises from the confinement of the order parameter fluctuations spectrum near the critical point ($T_\lambda$)~\cite{Fisher1978}. We show in Fig.~\ref{Fig_S7} the magnitude of these forces driving the superfluid nanomechanical resonator as a function of the gap between the electrodes. By lowering the drive voltage, we decrease the magnitude of the electrostatic and electrostrictive forces with respect to the Casimir forces. The electromagnetic Casimir force is nearly independent of temperature near $T_\lambda$ but the critical Casimir force is strongly temperature dependent and is nonzero only near $T_\lambda$ where its magnitude is given by Eq.~\ref{eq.critcas}. It may be possible to realize a direct measurement of the critical Casimir force in liquid \he4 near $T_\lambda$ by measuring the deflection of the plates if their stiffness is small enough. For example, using a standard capacitance bridge one can measure the capacitance with a resolution of $\delta C/C=10^{-8}$, which means a resolution on the deflection of the same amount $x/h_{\rm eff}=10^{-8}$. The force sensitivity of our device is given by $F_{\rm min}=k_{\rm plate}h_{\rm eff}10^{-8}\simeq180\ \mathrm{nN}$.

This force sensitivity can be increased by reducing the gap between the electrodes and reducing the plate stiffness. As an example, with $h_{\rm eff}\sim100$ nm and the same stiffness $k_{\rm plate}$, one can already have a force sensitivity ($F_0\sim10$ nN) high enough to probe the critical Casimir force ($F_{\rm Cas}^{crit}\sim1\ \mu$N).

An exhaustive summary of forces should include the dissipative and reactive forces induced by the thermal effects in superfluid \he4 flows, as mentioned by Backhaus~\etal~\cite{Backhaus1997a,Backhaus1997b}

\bibliography{apssamp}

\end{document}